\documentclass[american,aps,prl,reprint,showpacs,superscriptaddress,altaffilletter]{revtex4-1}
\usepackage{lmodern}
\usepackage{lmodern}
\usepackage[T1]{fontenc}
\usepackage[utf8]{inputenc}
\usepackage{color}
\usepackage{babel}
\usepackage{textcomp}
\usepackage{amsmath}
\usepackage{amssymb}
\usepackage{graphicx}
\usepackage[unicode=true,
 bookmarks=true,bookmarksnumbered=false,bookmarksopen=false,
 breaklinks=false,pdfborder={0 0 0},pdfborderstyle={},backref=false,colorlinks=true]
 {hyperref}
\hypersetup{
 pdfborderstyle={},pdfborderstyle={},pdfborderstyle={},pdfborderstyle={},pdfborderstyle={},pdfborderstyle={},linkcolor=blue,citecolor=blue}

\makeatletter
\usepackage{comment}

\usepackage[separate-uncertainty=true, group-separator =]{siunitx}

\makeatother

\begin{document}
\title{Faraday-waves contact-line shear gradient induces streaming flow and
tracers' self-organization: from rotating rings to spiral galaxy-like
patterns}
\author{H\'ector Alarc\'on}
\email{hector.alarcon@uoh.cl}

\affiliation{Instituto de Ciencias de la Ingenier\'ia, Universidad de O'Higgins,
Av. Libertador Bernardo O'Higgins 611, Rancagua, Chile}
\affiliation{Departamento de F\'isica, Facultad de Ciencias F\'isicas y Matem\'aticas,
Universidad de Chile, Av. Blanco Encalada 2008, Santiago, Chile}
\author{Matías Herrera}
\affiliation{Departamento de F\'isica, Universidad de Santiago de Chile,
 Av. Ecuador 3493, Estaci\'on Central, Santiago, Chile}
\author{Nicolas Périnet}
\affiliation{Departamento de F\'isica, Facultad de Ciencias F\'isicas y Matem\'aticas,
Universidad de Chile, Av. Blanco Encalada 2008, Santiago, Chile}
\author{Nicolás Mujica}
\affiliation{Departamento de F\'isica, Facultad de Ciencias F\'isicas y Matem\'aticas,
Universidad de Chile, Av. Blanco Encalada 2008, Santiago, Chile}
\author{Pablo Gutiérrez}
\email{pablo.gutierrez@uoh.cl}
\affiliation{Instituto de Ciencias de la Ingenier\'ia, Universidad de O'Higgins,
Av. Libertador Bernardo O'Higgins 611, Rancagua, Chile}
\author{Leonardo Gordillo}
\email{leonardo.gordillo@usach.cl}
\affiliation{Departamento de F\'isica, Universidad de Santiago de Chile,
 Av. Ecuador 3493, Estaci\'on Central, Santiago, Chile}
 
\begin{abstract}
In this letter we experimentally demonstrate self-organization of
small tracers under the action of longitudinal Faraday waves in a
narrow container. We observe a steady current formation dividing the
interface in small cells given by the symmetries of the Faraday wave.
These streaming currents are rotating in each cell and their circulation
increases with wave amplitude. This streaming flow drives the tracers
to form patterns, whose shapes depend on the Faraday wave's amplitude:
from low to high amplitudes we find dispersed tracers, a narrow rotating
ring and a spiral galaxy-like pattern. We first describe the main
pattern features, and characterize the wave and tracers' motion. We
then show experimentally that the main source of the streaming flow
comes from the time and spatial dependent shear at the wall contact
line, created by the Faraday wave itself. We end by presenting a 2D
model that considers the minimal ingredients present in the Faraday
experiment, namely the stationary circulation, the stretching component
due to the oscillatory wave and a steady converging field, which combined
produce the observed self--organized patterns. \\

PACS: 89.75.Kd: Pattern formation in complex systems; 47.55.np Contact
lines; 47.35.Bb Gravity waves
\end{abstract}
\maketitle
The free surface of liquids is usually covered by particles. In the
ocean these can include impurities, living organisms, nutrients, seeds,
garbage or even bubbles. Due to ocean's continuous motion, these \emph{tracers}
are constantly arranging themselves as they respond to sea waving
\citep{Haller_2015}. In more controlled, laboratory conditions, there
are several mechanisms that can influence tracers' assembly into patterns.
Formation of patterns has been attributed to particles properties,
as wetting \citep{Falkovich:2005ei,Sanl:2014hs,Dalbe:2011ul,Vella_2015},
filling fraction of the surface \citep{Berhanu:2010gm,Sanl:2014hs},
or inertial effects \citep{Santamaria:2013gv}. But tracers can also
be steered by underlying flows \citep{Boffetta:2004bc,Cressman:2004dc,GutierrezAumaitre_2016b,Lovecchio:2013cd}.
Even in controlled settings as standing Faraday waves, several mechanisms
lead to fluid motion decoupled from the periodicity of the waves \citep{Francois:2014is,Filatov:2016jc},
where tracers' motion becomes difficult to predict. An example of
the difficulties faced when predicting motion is the simple case of
propagating waves generated by an oscillating plunger, which has shown
a surprising flow reversal when crosswise waves form \citep{Punzmann:2014es}.
In particular, walls play a relevant role because dynamic wetting
between liquid and moving walls determines flow boundary conditions
\citep{TingPerlin_JFM1995,GordilloMujica_2014,Perinet_JFM2017,Huang:2020er}
and strongly influences dissipation \citep{JiangPerlin_PoF2004,Gordillo:2014gl,Dollet:2020dn}.

In this letter, we identify a new mechanism leading to streaming flows
on standing Faraday waves: a space-time dependent shear, produced
at the oscillatory liquid-wall contact line. The steadiness and localization
of the induced streaming flow, drives surface tracers to an exotic
self--organization, documented and explained here for the first time.
\begin{figure}[!b]
\begin{centering}
\includegraphics[width=1\columnwidth]{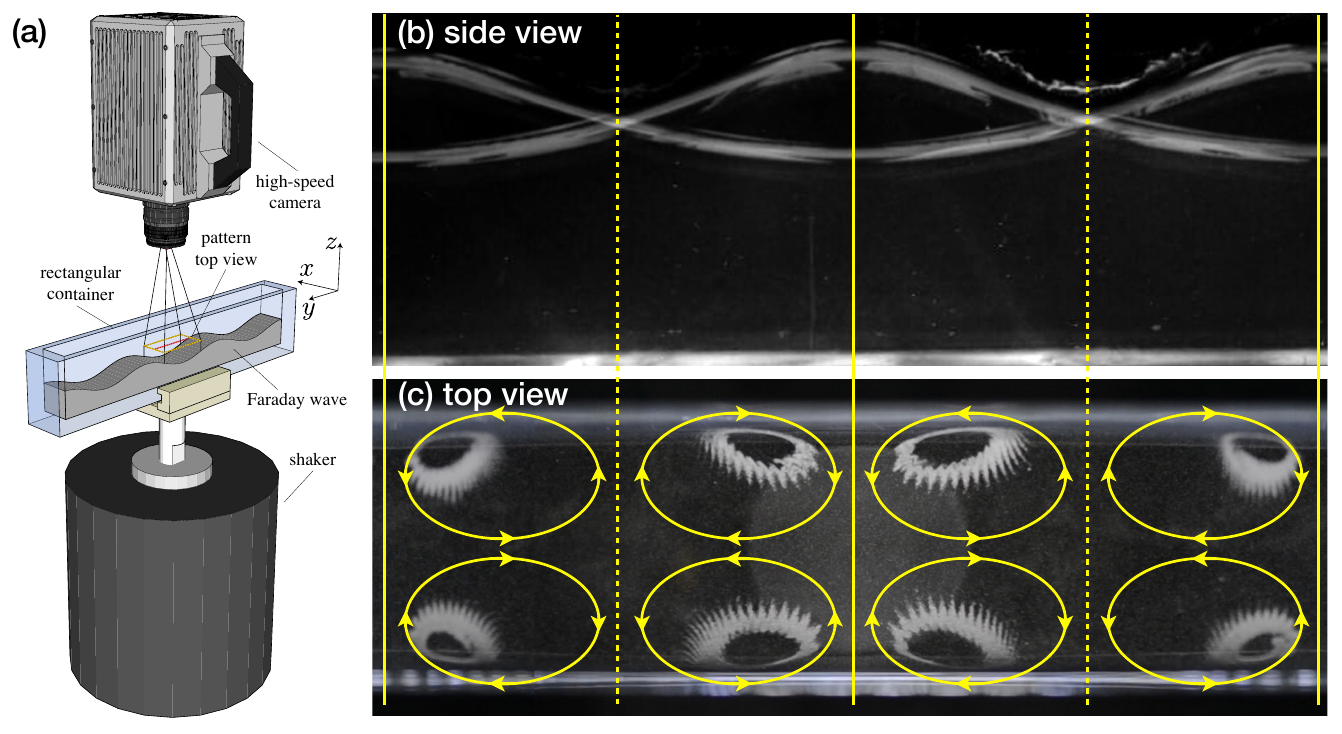}
\par\end{centering}
\caption{{\small{}Experimental setup and }spiral galaxy-like patterns{\small{}.
(a) An }electromechanical shaker vibrates vertically the acrylic container
with a water-surfactant liquid mixture and a small amount of silver-coated
hollow glass micro-spheres. Side (b) and top (c) views of the patterns
observed on a single Faraday wavelength. Solid (dashed) lines indicate
anti-nodes (nodes). The observed streaming flow is sketched as arrowed
ovals in (c).\label{fig:setup}}
\end{figure}

\begin{figure}[!t]
\begin{centering}
\includegraphics[width=1\columnwidth]{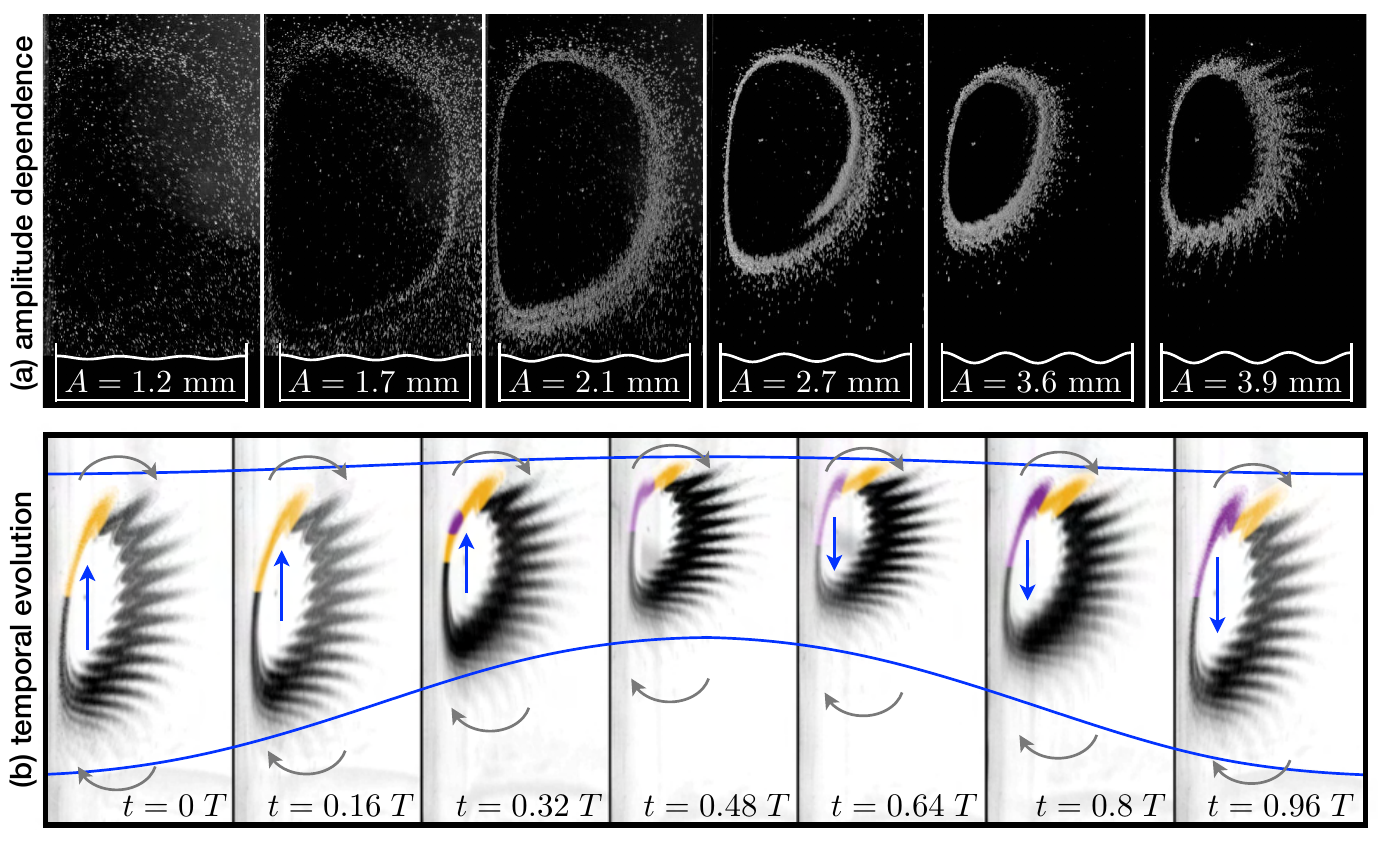}
\par\end{centering}
\caption{Elementary spiral galaxy-like pattern. (a) Pattern at a given phase
for various wave amplitudes $A$. (b) Breathing motion during a single
oscillation (gray scale was inverted for clarity); the formation of
a new arm is highlighted in purple. \label{fig:galaxies}}
\end{figure}

This letter is organized as follows: we start by presenting the Faraday--waves
setup, where we observe tracer self--organized patterns, which we
describe. We continue by focusing in measurements of the streaming
flow, which is, as we hypothesize, produced by a shear gradient on
the contact line. We replicate this flow with a simplified setup that
only engenders shearing of the liquid at the solid wall. We finish
by presenting a 2D model that considers the minimal ingredients present
in the Faraday experiment, which reproduces the observed patterns.

\paragraph{Experiment on Faraday waves.\label{sec:exp_faraday}}

The experimental setup used to produce Faraday waves is a rectangular
acrylic container with interior dimensions of $26.5\times280\times55$
\SI{}{\milli\metre\tothe{3}} attached to an electromechanical shaker
allowing vertical vibrations {[}Fig.\ref{fig:setup}(a){]}. The container
is filled up to \SI{30}{\milli\metre} deep with an aqueous solution
of \SI{218}{\gram} of distilled water and \SI{2}{\milli\litre} of
Kodak Photo-Flo, used as wetting agent \citep{Wu_prl1984}, and sprinkled
homogeneously with \SI{0.1}{\milli\gram} of \SI{10}{\micro\metre}
diameter silver-coated hollow glass micro-spheres, whose mass density
is $\rho=\SI{1.4}{\gram\centi\metre\tothe{-3}}$ (Dantec Dynamics,
S-HGS). The container vibrates sinusoidally $z(t)=a\sin\left(\omega_{d}t\right)$
, at the driving amplitude $a$ and the fixed frequency $f_d=\omega_d/2\pi=\SI{8.30}{\hertz}$.
This driving induces super-critical (non-hysteretic) Faraday waves
at $f = f_d/2 = \SI{4.15}{\hertz}$, above a normalized critical acceleration
amplitude $\Gamma_{c}=a_{c}\omega_{d}^{2}/g=0.144\pm0.001$  ($g$
is the gravitational acceleration). Under these experimental conditions,
the observed stationary Faraday wave forms a $(3,0)$-mode, i.e. three
wavelengths in the longwise direction with no crosswise component.

To characterize the flow due to Faraday waves, we measured simultaneously
the position of the tracers and the amplitude of the waves. In order
to visualize and track the tracers, we set a camera above the container
and a focused razing horizontal light from the side. The recording
frequency is $100$ fps with a $1920\times1080$ $\mathrm{pix}^{2}$
resolution. The local height of Faraday waves is simultaneously measured
from a side view using a tilted mirror {[}Fig.~\ref{fig:setup}(b){]}.

\paragraph{Flow structure and tracers' self-organization.\label{sec:vel_patt_mod}}

As soon as Faraday waves emerge, tracers begin to circulate inside
cells limited in the $x$--direction by the central plane of the
container and the walls, and in the $y$--direction by node and anti-node
planes, as indicated in Fig.~\ref{fig:setup}(c). The circulation
is steady and its origin will be discussed later. Cells adjacent to
antinodes or center lines shelter identical but counter-rotating patterns,
as shown in Fig.~\ref{fig:setup}(c). Near walls, tracers always
go from node to antinode \footnote{See Supplemental Material at ... for experimental videos, technical
details and numerical simulations of the advection model\label{fn:Supp}}. This circulation--cell structure is compatible with previous streaming
flow observations \citep{Perinet_JFM2017}.

Tracers display four organization regimes depending on the Faraday
wave amplitude $A$, as shown in Fig. \ref{fig:galaxies}(a) and \footnotemark[1]\ref{fn:Supp}:
1) for $A\leq\SI{1.5}{\milli\metre}$ the rotating tracers spread
more or less homogeneously over each cell, 2) above $A\approx\SI{1.5}{\milli\metre}$,
tracers in each cell shrink into a rotating ring, 3) for $A>\SI{2.8}{\milli\metre}$,
the rotating ring sharpens and exhibits \emph{arms} in a shape that
is reminiscent of spiral galaxies. The arms form after each wave cycle
and their length increases with $A$. 4) Finally, for $A>\SI{4.3}{\milli\metre}$,
arms become very long and oscillate in a time scale much larger than
the one of the Faraday waves (not shown in Fig.~\ref{fig:galaxies}).

In streaming galaxy-like patterns, one new arm emerges in each oscillation
period. The arm formation can be observed along a representative cycle
shown in Fig.~\ref{fig:galaxies}(b). In one hand, the pattern compresses
and stretches periodically, as a slave mode of the free surface oscillation.
This induces a back-and-forth motion during the oscillation period,
as schematized with vertical blue arrows in Fig. \ref{fig:galaxies}(b).
On the other hand, the steady streaming slowly rotates the pattern,
as shown with the curved arrows in the same figure. Arms appear in
the stage of surface compression: tracers near the wall (emphasized
in purple) are traveling faster than their counterparts near the anti-node
(in yellow). Therefore, fluid parcels fold, giving birth to an arm.

In order to evaluate the role of the particle nature of tracers in
the pattern formation, we ran the experiment using soluble dyes (ink)
and the patterns remained. Moreover, increasing the concentration
of tracers by a factor 5 keeps the observed patterns unaltered. These
observations demonstrate that the patterns are generated by the flow
itself, and tracers interaction or buoyancy are not relevant for their
emergence, contrary to what was observed in previous experiments with
larger particles \citep{Falkovich:2005ei,Sanl:2014hs}.

\paragraph{Steady Streaming.\label{sec:vel_patt}}

In our Faraday experiments we observe a steady circulation, as schematized
in Fig.\ref{fig:setup}(c). Due to their small size, tracers follow
fluid parcels. To characterize their displacement, we use image cross--correlation.
However, due to high wave amplitudes, large in--plane deformation
limits global measurements. We specifically refer to the interface
stretching that occurs due to the longitudinal wave. 
\begin{figure}[!t]
\begin{centering}
\includegraphics[width=0.95\columnwidth]{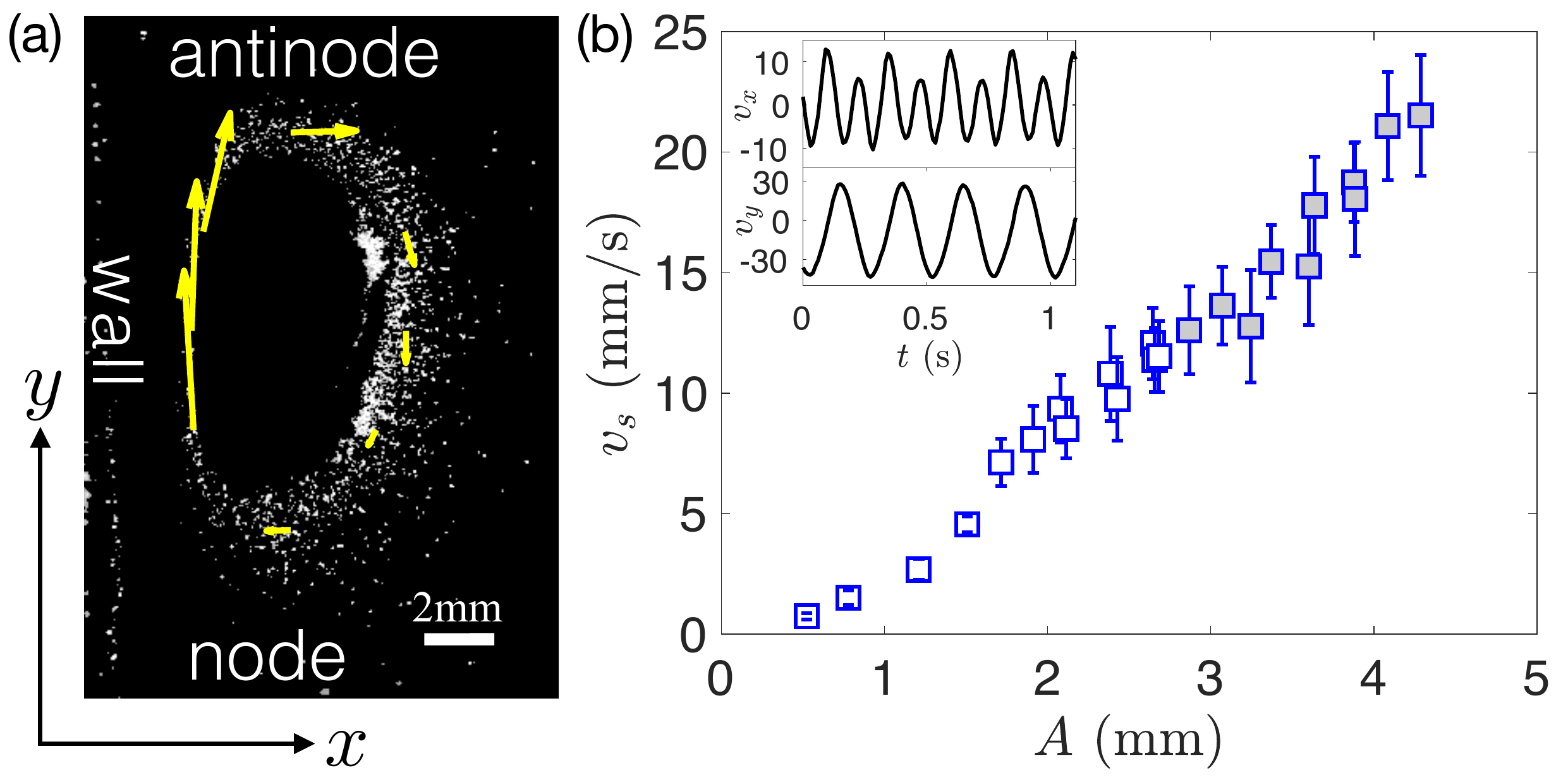}
\par\end{centering}
\caption{(a) Tracer velocities around a unit-cell rotating ring, for $A=\SI{2.1}{\milli\metre}$.
Streaming velocities, shown as yellow arrows, are measured at 8 different
locations around the ring, being much larger close to the wall. (b)
Streaming speed averaged at the three points near the wall as a function
of the wave amplitude. Solid points correspond to measurements with
the arms well developed in the circulating ring. Inset: examples of
instantaneous velocity components at a given point around the ring.\label{fig:circulation}}
\end{figure}
 Therefore, we focused on 8 small sections of the ring formed by tracers,
shown in Fig.~\ref{fig:circulation}(a). We track the longitudinal
in-plane displacement due to the Faraday wave as in a Lagrangian frame
of reference. The velocity of this frame of reference is time dependent
and obtained by linear interpolation along $y$ between the velocities
of the two extremities of the ring. Then we compute image cross--correlation
between small windows that follow this frame of reference. The total
velocity is then deduced from the prescribed Lagrangian displacement
plus the one measured from image cross-correlation. Figure \ref{fig:circulation}(a)
shows an instantaneous picture, and the inset of Fig. \ref{fig:circulation}(b)
shows the time evolution of the $x$-and-$y$-velocity components
of a single point near the wall.

Temporal evolution in $v_{y}$ shows a sub--harmonic oscillation,
while $v_{x}$ exhibits a harmonic mode. While the Faraday wave is
the dominant motion along $y$, the $x$-direction is more sensitive
to meniscus waves. Both velocity components have a well--defined
non--zero time--average, corresponding to the steady streaming.
To evaluate this magnitude in function of $A$, we define $v_{s}$
as the average speed between the three points near the wall. The main
panel of Fig. \ref{fig:circulation}(b) shows that $v_{s}$ increases
first quadratically, then nearly linearly with $A$.

The final, and most remarkable feature of the streaming flow is its
heterogeneity along the ring. Indeed, Fig. \ref{fig:circulation}(a)
shows that the velocity is around $6$ times larger near the walls
than close to the container's center line. Therefore, near the wall,
the liquid experiences a large acceleration linked to a strong shear
gradient. For standing waves, due to the dynamical variation of the
surface elevation between nodes and antinodes, a space-time dependent
shear gradient is ubiquitous whenever walls are present. Are circulating
streaming flows a generic feature when a space-time variable shear
gradient is exerted on a liquid-solid-air contact line? We will answer
this by isolating the effect of the oscillatory shear gradient.

\paragraph{Rotating disk experiment.\label{sec:disk}}

\begin{figure}[!t]
\begin{centering}
\includegraphics[width=0.95\columnwidth]{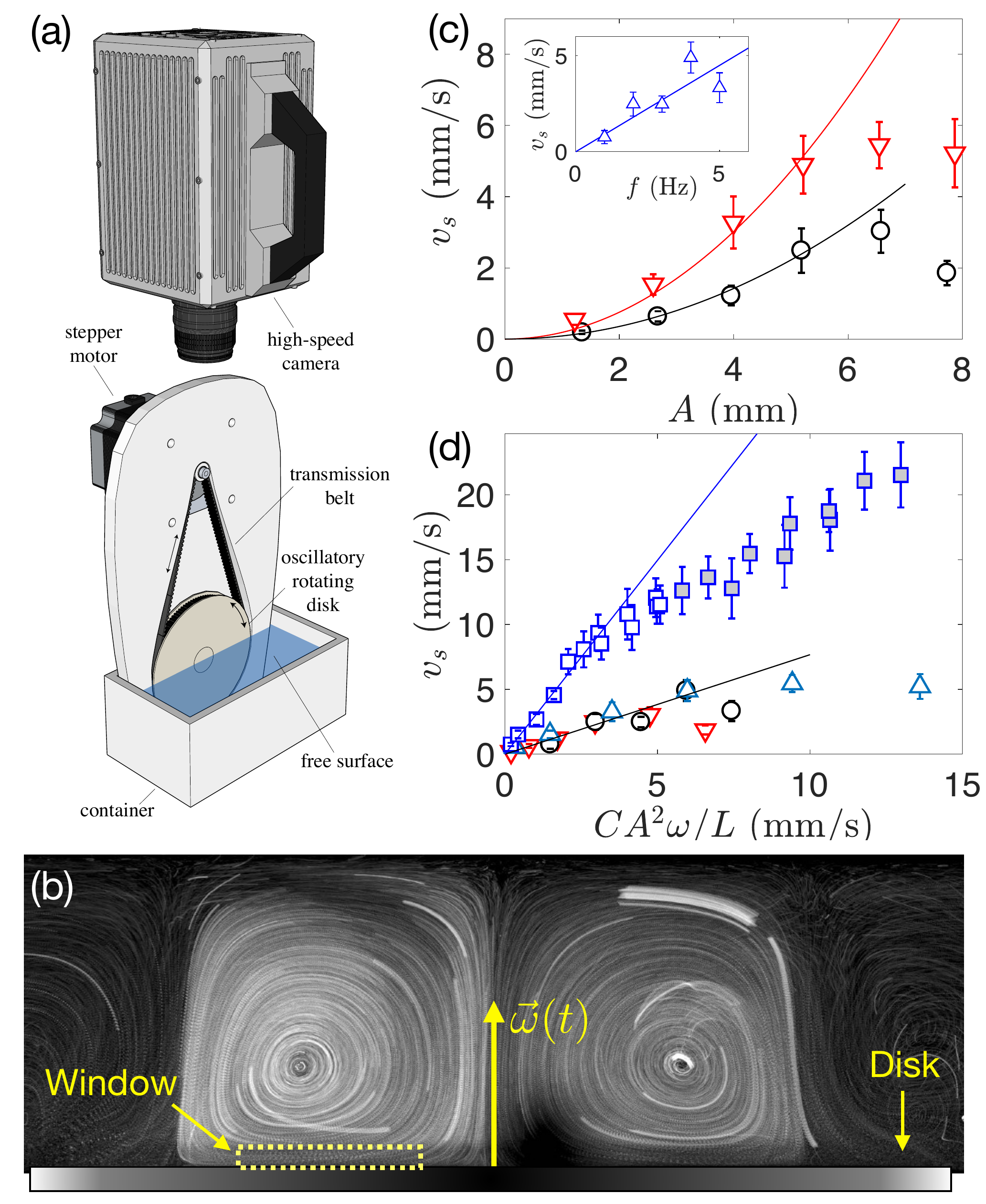}
\par\end{centering}
\caption{(a) Schematic illustration of the rotating disk setup. (b) Top view
of the observed streaming pattern, $A=R\theta=\SI{2.67}{\milli\metre}$
and $f = \SI{2}{\hertz}$. The PTV measurement window is shown as
dashed rectangle. (c) Streaming speed as function of disk oscillation
amplitude $A$, for $f = \SI{2}{\hertz}$ $(\triangledown)$ and $f = \SI{4}{\hertz}$
$(\bigcirc)$. Inset: $v_{s}$ linear frequency dependence for fixed
$A$. (d) $v_{s}$ as function of $C\omega A^{2}/L$, with $L=R$
and $C=0.35$ for the disk, and $L=\lambda$ and $C=2.5$ for Faraday
data. The disk data collapses on a single curve, showing a saturation
at higher values. For both experiments an $A^{2}$ scaling is observed
for the lower amplitude regime, shown as straight lines as guides
to the eye. \label{fig:disk}}
\end{figure}

A circular disk (of radius $R=40$ mm) rotates periodically between
angles $\theta$ and $-\theta$, driven by a stepper motor (see Fig.
\ref{fig:disk}(a)). The disk is partially immersed in water up to
its center, so that any point oscillates with an amplitude $r\theta$,
with $r$ the distance from the center (see Fig. \ref{fig:disk}(a)),
and with a maximum arc amplitude $A=R\theta$. We limited $A$ to
8 mm in order to avoid sloshing motion. Oscillation frequency $f$
is set independently. The disk is made of acrylic, having similar
wetting properties as the container used for Faraday waves. The disk
oscillatory motion, therefore, simulates the spatio-temporally varying
triple-line position distinctive of Faraday waves.

Fig. \ref{fig:disk}(b) presents a digitally overexposed image, revealing
flow streamlines. The result is striking, as it shows a steady circulation
reminiscent of the one observed in the Faraday experiment for low
amplitudes. In order to further characterize this steady streaming,
we measured the spatial average streaming speed $v_{s}$ in a window
close to the disk, as shown in this figure. Here, because motion is
much slower than in the Faraday experiment, we are able to perform
simple Particle Tracking Velocimetry (PTV) measurements on tracers
that are detected. The streaming speed is then computed as a time-spatial
average from individual tracer measurements. We performed measurements
for fixed $f$ increasing $A$, and also for fixed $A$ increasing
$f$. We observe a linear increase of $v_{s}$ with $f$, and a quadratic
dependence on $A$, with a saturation at the highest values {[}Fig.
\ref{fig:disk}(c){]}. When plotting $v_{s}$ in function of $A^{2}\omega/R$,
all disk measurements collapse on a single curve (see Fig. \ref{fig:disk}(d)).
The consistency of both scaling laws at low amplitudes allows us to
conclude that the circulation comes from the shear gradient of vertical
velocity at the walls, similar to what was proposed in \citep{Nicolas:2009ki,Perinet_JFM2017},
but specifically applied to the contact line. Indeed, the data is
consistent with $v_{s}\propto~\langle w_{\sim}\partial_{y}w_{\sim~}\rangle/(2\pi f)$,
where the vertical velocity component of the contact line at the wall
is $w_{\sim}=2\pi fA\sin(ky)\cos(2\pi ft)$ for Faraday waves and
$w_{\sim}=2\pi fAy\cos(2\pi ft)/R$ for the disk. Here, the brackets
$\langle\,\,\rangle$ stand for the spatial and temporal average in
each case. The constant $C$ used in Fig. \ref{fig:disk}(d) is the
result of the numerical factors for each case, considering also the
spatial averaging. The law $v_{s}\propto~\langle w_{\sim}\partial_{y}w_{\sim~}\rangle/(2\pi f)$
also gives an explanation to the generation of eddies on the crosswise
waves in \citep{Punzmann:2014es}, which share the common feature
of sheared oscillating contact lines. It also consistently predicts
that nodes/antinodes generate inward/outward jet streams in such configuration.

\paragraph{Simple advection model.\label{sec:disk-1}}

The experiment with the oscillatory rotating disk successfully isolates
the gradient of shear exerted on the contact line as it oscillates
on the wall. It clearly shows that the primary origin of the circulation
of the observed patterns in Faraday waves is not the presence of the
wave itself, but the shear gradient along the contact line. The question
that naturally arises is whether it is possible to go further with
a model that is not only able to predict a rotating velocity field
but also complex galaxy-like structures.

\begin{figure}[!t]
\begin{centering}
\includegraphics[width=0.9\columnwidth]{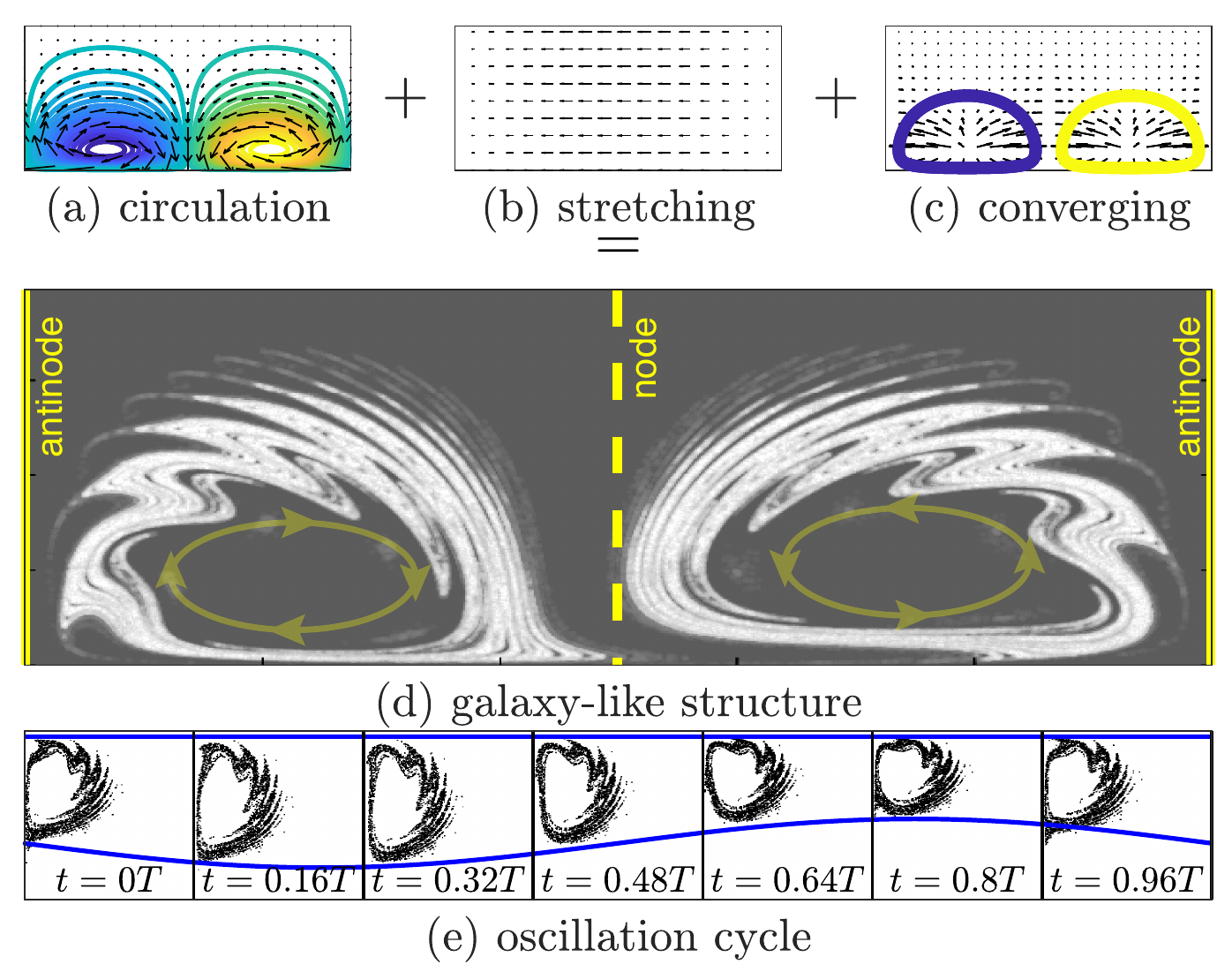}
\par\end{centering}
\caption{(a) Steady circulation, (b) periodic stretching and (c) a steady converging
field featured by a ring, are the basic ingredients to generate galaxy-like
structures. The superposition of the velocity fields allows agglomeration
of particles (d) and similar dynamics, including rotation, breathing
and formation of arms in every cycle (e) as those that emerge in patterns
on Faraday waves. \label{fig:model}}
\end{figure}

We have concluded that there are three fundamental features required
for the formation of galaxy-like structures, which can be put into
a simple 2D phenomenological advection model. These are:

\textit{1. A steady circulation. }To build a velocity field that has
the features observed in Faraday cells, we consider a modification
of the classic flow of a 2D rectangular cavity driven by a moving
wall. Instead of imposing a uniform velocity at one of the boundaries
of the cavity, we imposed a non-uniform velocity, $\propto-\sin\left(2ky\right)$
in the lower boundary at $x=0$, to mimic the effect of the sheared
contact line. Free-slip boundary conditions are applied on the rest
of the boundaries as they correspond to the symmetry planes of the
Faraday-wave cells. The streamlines due to the presence of this flow
can be calculated analytically \footnotemark[1]\ref{fn:Supp} and
are shown in Fig.~\ref{fig:model}(a).

\textit{2. A periodic stretching}. Figure~\ref{fig:galaxies} shows
that the structures undergo periodic stretching in time. To account
for this, we added an oscillatory field given by $v\left(x,y,t\right)=-\epsilon\sin ky\cos\omega t$,
where $\epsilon$ is the amplitude of the deformation due to the main
motion of the stationary Faraday wave as shown in Fig.~\ref{fig:model}(b).

\textit{3. A steady converging field.} This flow accumulates the particles
into a ring (see Fig.~\ref{fig:model}(c)). To account for this,
we took the rotating velocity field and manipulated it to generate
a secondary flow that converges particles into one of its streamlines
\footnotemark[1]\ref{fn:Supp}.

We ran simple numerical simulations in which an initially random distribution
of particles is advected by the superposition of these three velocity
fields. As shown in \footnotemark[1]\ref{fn:Supp}, particles agglomerate
in structures that are incredibly similar to those observed on Faraday
waves as shown in Fig.~\ref{fig:model}(d). The model captures successfully
very specific and complex features of the motion of the structures:
the breathing dynamics {[}Fig.~\ref{fig:model}(e){]}, the mechanism
for generation of arms in each period and a gradual decrease of the
length of the arms as they approach the wall.

In conclusion, we have found experimentally that the flow at the interface
of Faraday waves undergoes a transition from circulating rings to
spiral-like galaxies as the wave amplitude increases. The observed
galaxy arms are created at each period due to the strong shear gradient
at the wall and the fast deceleration of the circulating ring into
the bulk of the interface, away from the walls. We demonstrated that
the main source of the streaming flow comes from the time and spatial
dependent shear on the contact line at the wall boundary, induced
by the Faraday wave itself. The streaming flow is quite generic, as
a similar circulation flow is observed in the oscillating rotating
disk experiment. Both experiments follow the scaling $v_{s}\propto~\langle w_{\sim}\partial_{y}w_{\sim~}\rangle/(2\pi f)$
for the average streaming speed near the wall, where $w_{\sim}$ is
the contact-line vertical velocity component on the wall. The proposed
scaling law provides also an explanation to the eddies and jet streams
documented in \citet{Punzmann:2014es}, setting the basis for the
understanding of flow reversal on propagating waves. Finally, we propose
a simplified 2D model that considers the minimal ingredients that
are necessary to obtain spiral-galaxy-like patterns, namely the stationary
circulation, the stretching component due to the oscillation and a
steady converging field. The model imitates the patterns observed
experimentally in many regards. As a perspective, this flow, which
is periodic with a nonlinearly generated stationary component, is
an interesting system for studying Lagrangian coherent structures
\citep{Haller_2015,FarazmandHaller_2012}, specially considering that
the stirring process ends with a complex pattern of tracers with a
high degree of spatiotemporal organization in a rather simple experimental
configuration.

\begin{acknowledgments} 
H. A. acknowledges the support of Fondecyt postdoctoral program, grant
3160341. L.G., P.G. and N.M. acknowledge support of ANID through the
Fondecyt grants Nº11170700, Nº11191106 and Nº1180636, respectively.\end{acknowledgments}

\bibliographystyle{apsrev4-1}

\end{document}